\documentclass[english,aps,preprint]{revtex4}
\usepackage[T1]{fontenc}
\usepackage[latin1]{inputenc}
\usepackage{amssymb}


\begin{document}

\preprint{BROWN-HET-1367}

\title{Ultra High Energy Cosmic Rays and de Sitter Vacua}

\author{Gian Luigi Alberghi}

\email{gigi@het.brown.edu}

\affiliation{\textsl{Dipartimento di Fisica, Universit\`{a} di Bologna and  I.N.F.N,
Sezione di Bologna, Bologna, Italy.}}

\author{Kevin Goldstein}

\email{kevin@het.brown.edu}

\affiliation{Physics Department, Brown University, Providence, R.I. 02912, USA}

\author{David A. Lowe}

\email{lowe@brown.edu}

\affiliation{Physics Department, Brown University, Providence, R.I. 02912, USA}

\begin{abstract}
The production of ultra high-energy cosmic rays in de Sitter invariant
vacuum states is considered. Assuming the present-day universe is
asymptoting toward a future de Sitter phase, we argue the observed
flux of cosmic rays places a bound on the parameter $\alpha$ that
characterizes these de Sitter invariant vacuum states, generalizing
earlier work of Starobinsky and Tkachev. If this bound is saturated,
we obtain a new top-down scenario for the production of super-GZK
cosmic rays. The observable predictions bear many similarities to
the previously studied scenario where super-GZK events are produced
by decay of galactic halo super-heavy dark matter particles. 
\end{abstract}
\maketitle

\section{introduction}

It has been known for a long time that de Sitter space possesses nontrivial
vacuum states that are invariant under the symmetries of the space
\cite{Chernikov:1968zm,Tagirov:1973vv,Allen:1985ux,Mottola:1985ar},
which we call the $\alpha$-vacua. Physical applications of these
states have recently been explored in the context of inflation, where
they can lead to potentially observable corrections to the spectrum
of cosmic microwave background fluctuations \cite{Easther:2001fi,Easther:2001fz,Easther:2002xe,Bergstrom:2002yd,Danielsson:2002kx,Danielsson:2002mb,Danielsson:2002qh,Goldstein:2002fc,Alberghi:2003am}.
If such states can be relevant during inflation, it is natural to
ask whether such states can have other observable consequences today.
An initial study along these lines was made in \cite{Starobinsky:2002rp}
where the contribution to the cosmic ray spectrum was considered. 

The cosmic ray spectrum has a feature at around $5\times10^{18}$eV
where the power-law spectrum flattens from $E^{-3.2}$ to $E^{-2.8}$
as $E$ increases, which suggests a transition from a galactic component
of conventional astrophysical origin, to a component of extra-galactic
origin. Some recent reviews of theoretical and experimental prospects
for the study of these ultra high energy cosmic rays may be found
in \cite{Anchordoqui,Bhattacharjee:1998qc}. Above about $10^{20}$eV,
protons rapidly lose energy due to their interaction with the cosmic
microwave background, leading to the GZK cutoff \cite{Greisen:1966jv,Zatsepin:1966jv}.
A handful of events above this bound have been observed, and there
is still come controversy over whether or not the cutoff has been
observed \cite{Takeda:1998ps,Hayashida:2000zr,Hires}. Extremely high
energy cosmic rays $\gtrsim10^{20}$eV are difficult to explain using
conventional physics because likely sources lie outside the $100$Mpc
range of $10^{20}$eV protons. A wide variety of scenarios have been
proposed to account for the super-GZK events, which break down into
two main classes: bottom-up mechanisms where charged particles are
accelerated in large scale magnetic fields, and top-down mechanisms
where exotic particles/topological defects produce extremely high
energy cosmic rays via decay.

The $\alpha$-vacua provide us with a new top-down mechanism for the
production of extremely high energy cosmic rays. As noted in \cite{Bousso:2001mw},
a comoving detector can detect transitions of arbitrarily large energies
(which we assume are cutoff near the Planck scale). The exception
is the Bunch-Davies vacuum, where a detector measures a thermal response
with a temperature of order the Hubble scale. These very high energy
transitions in a generic $\alpha$-vacuum can then account for production
of extremely high energy cosmic rays.

Starobinsky and Tkachev \cite{Starobinsky:2002rp} argued that if
the $\alpha$-vacua %
\footnote{Starobinsky and Tkachev analyze the situation where the vacuum is
a mode number independent Bogoliubov transformation relative to the
Bunch-Davies vacuum. Although they did not refer to them as such,
these are the $\alpha$-vacuum states studied earlier in \cite{Chernikov:1968zm,Tagirov:1973vv,Mottola:1985ar,Allen:1985ux}.%
} do contribute to the ultra high energy cosmic ray spectrum at around
$10^{20}$eV, then the value of $\alpha$ becomes so tightly constrained
that it would not produce observable effects during inflation. In
the present paper, we revisit this question and argue a much weaker
constraint on $\alpha$ follows from cosmic ray observations. Our
analysis also has bearing on the general question of what a low-energy
observer will see in an $\alpha$-vacua. The upshot of our analysis
is that because production of cosmic ray flux necessarily violates
de Sitter invariance, the production rate will be proportional to
the background number density of matter, which leads to a much suppressed
production rate versus the estimates of \cite{Starobinsky:2002rp}.
This rate is calculated in detail in section \ref{sec:Unruh}. 

From this result we infer bounds on $\alpha$ from cosmic ray observations.
Assuming these bounds are saturated, we find the $\alpha$-vacua give
predictions very similar to extremely high energy cosmic ray production
via decaying super-heavy dark matter in the galactic halo. This scenario
has already been much studied in the literature \cite{Kuzmin:1998cm,Birkel:1998nx,Berezinsky:1998qv}.
We check that observable signals are out of reach in current neutrino/proton
decay detectors. Finally we argue CPT violation in an $\alpha$ vacuum
does not give rise to baryogenesis.

\section{Comoving Detector in de Sitter Space\label{sec:Unruh}}

In a de Sitter invariant vacuum state, all correlators are invariant
under the continuously connected symmetries of de Sitter space. In
particular, this implies that $\left\langle n^{\mu}\right\rangle =0$
for all 4-vectors $n^{\mu}$, such as the number flux. Equivalently,
the stress energy tensor in the de Sitter vacuum is proportional to
the metric $T_{\mu\nu}\propto g_{\mu\nu}$. Since the metric is diagonal
in comoving coordinates, this implies the absence of fluxes of energy
or momentum. However a comoving detector nevertheless makes transitions
due to its passage through the background spacetime, via the Unruh
effect. We will model the injection spectrum of ultra high energy
cosmic rays by viewing the universe today as de Sitter with $H=H_{0}$,
the value of the Hubble parameter today. We treat the background density
of ordinary Standard model matter as a small perturbation that explicitly
breaks the de Sitter symmetry. Under certain circumstances, we can
then treat these matter particles as Unruh detectors, which make transitions
to highly excited states via interaction with the nontrivial vacuum
state. 

The $\alpha$ parameter in principle can depend on the species of
field \cite{Goldstein:2003ut}, which introduces a high degree of
model dependence in the predictions %
\footnote{A number of works have pointed out potential problems with the $\alpha$-vacuum
scenario \cite{Kaloper:2002cs,Banks:2002nv,Einhorn:2002nu,Einhorn:2003xb,Collins:2003zv}.
These criticisms have been addressed in \cite{Danielsson:2002mb,Goldstein:2003ut,lowe3}.
In particular, in \cite{lowe3} we show with a particular definition
of the $\alpha$-vacua in interacting scalar field theory, the theory
is well-defined in a fixed de Sitter background. However when this
theory is coupled to conventional gravity, problems with locality
arise. Possible loopholes to this argument include: strongly coupled
physics at the cutoff scale that render the perturbative arguments
of \cite{lowe3} invalid; quantization on elliptic de Sitter \cite{Parikh:2002py},
where one gives up time orientability; or other exotic non-local trans-Planckian
effects that again fall outside the analysis of \cite{lowe3}. In
the present paper we only use properties of the free two-point function,
and we assume some loophole to the arguments of \cite{lowe3} is in
effect.%
}. For simplicity let us model the fields of observable matter by a
single scalar field $\chi$ and assume that a different field $\phi$
(for example, the inflaton) is in a nontrivial $\alpha$-vacuum, with
coupling $\chi^{2}\phi$. We assume an order $1$ coupling of $\phi$
to observable Standard model matter fields. The linear coupling of
$\phi$ then allows us to treat the $\chi$ particles as an Unruh
detectors (see \cite{Birrell:1982ix} for a review). 

As shown in \cite{Bousso:2001mw} the rate at which an Unruh detector
makes transitions is \begin{equation}
\Gamma=N_{\alpha}^{2}\left|1+e^{\alpha+\pi\Delta E/H}\right|^{2}\Gamma_{0}\label{eq:genform}\end{equation}
 where $\Gamma_{0}$ is the result in the standard Bunch-Davies vacuum,
$N_{\alpha}^{2}=\frac{1}{1-e^{\alpha+\bar{\alpha}}}$. $\Gamma_{0}$
is Boltzmann suppressed by a $e^{-2\pi\Delta E/H}$ factor, so for
large $\Delta E$, $\Gamma$ is proportional to $N_{\alpha}^{2}\left|e^{\alpha}\right|^{2}$
times a power of $\Delta E$. The injection spectrum is dominated
by $\Delta E\sim M_{c}$ the field theory cutoff scale, which we have
in mind to be of order the GUT scale $10^{16}$GeV. When we integrate
over $\Delta E$, dimensional analysis then implies the total transition
rate is\begin{equation}
\Gamma\approx N_{\alpha}^{2}\left|e^{\alpha}\right|^{2}M_{c}.\label{eq:rate}\end{equation}
Here we have assumed $e^{\alpha}$ is not so small that the $1$ dominates
in the $1+e^{\alpha+\pi\Delta E/H}$ factor of (\ref{eq:genform}).
It is straightforward to generalize this expression to models with
different couplings of $\alpha$-vacuum species to observable matter,
using the general formula (\ref{eq:genform}).

\section{Ultra High Energy Cosmic Ray Production}

Let us begin by reviewing the $\alpha$-vacuum scenario, as described
in \cite{Goldstein:2002fc}. During inflation, trans-Planckian effects
\cite{Martin:2000xs} can lead to a de Sitter invariant state that
differs from the conventional Bunch-Davies vacuum. If we invoke {}``locally
Lorentzian'' boundary conditions on modes, as described in \cite{Danielsson:2002kx,Danielsson:2002qh},
one finds \begin{equation}
e^{\alpha}\sim H/M_{c}.\label{eq:alnat}\end{equation}
 This modifies inflationary predictions for the cosmic microwave background
spectrum \cite{Danielsson:2002kx,Danielsson:2002qh,Goldstein:2002fc}.
At the end of inflation, the value of the cosmological constant changes
drastically. The squeezed state corresponding to the $\alpha$-vacuum
will then generate particles, producing an energy density of order
\cite{Goldstein:2002fc,Starobinsky:2002rp}\begin{equation}
\varepsilon\sim N_{\alpha}^{2}\left|e^{\alpha}\right|^{2}M_{c}^{4}.\label{eq:enden}\end{equation}
Provided $M_{c}\ll M_{Planck}$ this particle production does not
overclose the universe, and instead can be thought of as some component
of particle production during reheating. This energy density will
decay in a time of order $1/M_{c}$ (up to coupling dependent factors),
as is typical of unstable particle production during reheating.

At much later epochs, it is still possible to have a residual $\alpha$-vacuum
present. If the universe asymptotes to a de Sitter universe with cosmological
constant determined by the present value of $H$, the arguments of
\cite{Danielsson:2002kx,Danielsson:2002qh} again apply and we can
expect $\alpha$ given by (\ref{eq:alnat}) \cite{Goldstein:2002fc,Danielsson:2002mb}.
One can then ask what phenomena observers today will see to indicate
the presence of the $\alpha$-vacuum. In \cite{Starobinsky:2002rp}
the assumption was made that (\ref{eq:enden}) will be present for
all times, and they used this to constrain $\alpha$ by matching with
observed ultra high energy cosmic ray production. This assumption
is equivalent to computing the energy density of the $\alpha$-vacuum
with respect to the Bunch-Davies vacuum, but gives the wrong result
if the future asymptotic vacuum state is the $\alpha$-vacuum. In
this case, as we described in section \ref{sec:Unruh}, no additional
particle creation will be present in the de Sitter phase, and instead
the particle production will be determined by (\ref{eq:rate}), where
we treat background matter as individual Unruh detectors. 

In reality, the present universe is far from a pure de Sitter phase.
The pure de Sitter estimate of the production rate nevertheless should
be a reasonable order of magnitude estimate of the present rate of
high energy particle production. Of course without a more detailed
model for the dynamics that governs $\alpha$ we cannot make more
precise statements. 

Let us proceed then to compute the rate of high energy particle production
in an $\alpha$-vacuum. By the arguments of section \ref{sec:Unruh},
we can then treat each Standard model particle as an Unruh detector,
so (\ref{eq:rate}) gives the rate of production per unit volume as

\begin{equation}
\frac{dn}{dt}=\Gamma n\label{eq:dnt}\end{equation}
where $n$ is the number density of Standard Model particles %
\footnote{The earlier calculation of \cite{Starobinsky:2002rp} obtained $dn/dt\sim|e^{\alpha}|^{2}HM_{c}^{3}$
(converting to our notation). Our result (\ref{eq:dnt}) is suppressed
by a factor of order $HM_{Planck}^{2}/m_{p}M_{c}^{2}$, where the
reader should recall $H$ is the Hubble parameter today, and we have
substituted the critical density for $n$.%
}. In the situation of interest here, this density will be of order
the baryon number density $n_{B}$ which is typically \[
\begin{array}{cccccc}
n_{B} & = & 10\,\textrm{m}^{-3} & \approx & H^{2}M_{Planck}^{2}/m_{p} & \textrm{critical density}\\
 & = & 10^{6}\,\textrm{m}^{-3} &  &  & \textrm{interstellar space}\end{array}\]
where $m_{p}$ is the proton mass. Plugging in numbers, we find the
dominant source of high energy cosmic rays will come from within our
own galaxy due to interaction of visible and dark matter with the
$\alpha$ vacuum. Many of the predictions will therefore be similar
to the class of top-down models for ultra high energy cosmic ray (UHECR)
production from decaying super-heavy dark matter particles in the
galactic halo. For a galactic halo of size $r_{halo}$ , we find the
flux received on earth will be of order\[
j\approx\Gamma n_{B}r_{halo}.\]
The experimental bounds coming from UHECR production gives $j\, E^{2}\approx10^{24}$
eV$^{\textrm{2}}$ m$^{\textrm{-2}}$ s$^{\textrm{-1}}$sr$^{\textrm{-1}}$
at $E\approx10^{20}$ eV. Assuming $r_{halo}\approx10^{5}$ light
years, this translates into a bound

\begin{equation}
|e^{\alpha}|\lesssim10^{-42}\left(\frac{10^{16}\textrm{GeV}}{M_{c}}\right)^{1/2}.\label{eq:expbound}\end{equation}
This is to be compared with the {}``natural value'' $e^{\alpha}\sim H/M_{c}\approx10^{-61}M_{Planck}/M_{c}$
which is much smaller. We conclude then if ultra high energy cosmic
ray production is to be accounted for by the $\alpha$ vacuum then
the value of $\alpha$ must be much larger than its natural value. 

It is interesting to ask if such a large value for $\alpha$ today
might have other observable consequences. Let us also estimate the
time needed for a neutrino style detector to see a nontrivial interaction
with the $\alpha$-vacuum. The interaction rate per baryon is (\ref{eq:rate})
(taking $e^{\alpha}=H/M_{c}$, and $M_{c}=10^{16}$GeV) \[
\Gamma=10^{-76}s^{-1}\]
 which is about $36$ orders of magnitude smaller than current bounds
on proton decay rate. For $\alpha$ saturating the bound (\ref{eq:expbound})
and $M_{c}=10^{16}$GeV, we instead get\[
\Gamma=10^{-44}s^{-1}\]
which is only 4 orders of magnitude smaller than the bounds on proton
decay. We conclude that even if $\alpha$ is so large as to account
for UHECR production, other means of direct detection will be difficult.

Finally, one might ask whether the present analysis has some impact
on the spectrum of primordial inflation fluctuations. During the inflationary
phase, the effect of the $\alpha$-vacua on the primordial spectrum
has been discussed in \cite{Danielsson:2002kx,Danielsson:2002qh,Goldstein:2002fc,Easther:2002xe},
where it was found the amplitude of the spectrum was modulated by
a factor of the form $1+\mathcal{O}(H/M_{c})$. The particle production
effects described in the present paper will be absent in empty de
Sitter, and we expect the effect will be a small correction to the
energy density (\ref{eq:enden}) in the context of slow-roll inflation.
Note we already constrain (\ref{eq:enden}) to be less than the vacuum
energy density during inflation \cite{Goldstein:2002fc}. Therefore
we expect the particle production effects described here will have
negligible impact on the spectrum of primordial fluctuations.

\section{Baryogenesis}

An interesting feature of the $\alpha$ vacua is that they violate
CPT symmetry when $\alpha$ is not a real number \cite{Allen:1985ux,Bousso:2001mw}.
This opens the possibility that the $\alpha$ vacua could be used
to explain baryogenesis. If the CPT violation gives rise to particle/anti-particle
mass differences, then baryogenesis could occur in thermal equilibrium,
and might be relevant during the reheating phase at the end of inflation. 

Greenberg \cite{Greenberg:2002uu} has argued that particle/anti-particle
mass differences are only possible in flat space, if one gives up
locality. We can apply these general results in the short wavelength
limit of $\alpha$ vacuum propagators. As shown in \cite{Goldstein:2003ut}
the interacting propagators give rise to local commutators in $\alpha$
vacua. In this limit, de Sitter symmetry becomes local Lorentz symmetry,
so Greenberg's result will carry over. We conclude that $\alpha$
vacua do not lead to this type of baryogenesis.

\section{Discussion and Conclusions}

We have explored some of the phenomenological consequences of the
present universe being in an $\alpha$-vacua. It seems the most promising
way to directly detect a residual value for $\alpha$ today is via
observations of ultra high energy cosmic rays. As we have mentioned,
many of the predictions will be similar to production of ultra high
energy cosmic rays via decaying super-heavy dark matter particles
in the galactic halo \cite{Kuzmin:1998cm,Birkel:1998nx,Berezinsky:1998qv}.
See \cite{Bhattacharjee:1998qc,Barbot:2002gt,Sigl:2002yk} for some
recent results, and more extensive references. Let us briefly discuss
some of the features and constraints on this production mechanism.

Galactic halo cosmic ray production avoids the GZK cutoff, because
the absorption length of ultra high energy protons is of order 100
Mpc. The cosmic rays typically do not have time to scatter before
they reach us, so the observed spectrum should reflect the fragmentation
function of the primary decay. This has been computed using Monte
Carlo calculation in \cite{Barbot:2002gt} including effects of supersymmetry.
Perhaps the main problem one encounters in matching this with observation
is that the fragmentation functions suggest the fraction of gamma
rays versus protons is too high versus the experimental bound  \cite{Ave:2000nd,Ave:2001xn,Shinozaki:2002ve}.
This bound should become much better established in the upcoming Pierre
Auger Observatory \cite{Cronin:1992ir}. It is possible gamma rays
lose energy more efficiently than protons over the scales of interest,
which would ameliorate this problem. Searches for ultra high energy
neutrinos should provide a more robust test of this scenario.

The observed arrival directions of UHECRs exhibit a high degree of
isotropy on large scales, but clustering on smaller scales. This can
be consistent with a clumpy distribution of dark matter in the galactic
halo. The $\alpha$ vacuum scenario predicts additional anisotropy
if the coupling to visible and dark matter is comparable, but these
couplings are not well-constrained given the current level of understanding.

The main goal of the present work was to investigate observational
constraints on a residual value of $\alpha$ today. These constraints
easily allow for the theoretically preferred value of $\left|e^{\alpha}\right|\sim H(t)/M_{c}$.
It is fascinating the $\alpha$ vacua may also lead to a possible
explanation of the spectrum of UHECRs.

\begin{acknowledgments}
We thank Robert Brandenberger for comments on the manuscript. D.L.
thanks Mark Trodden for a helpful conversation. D.L. thanks the Departamento
de F\'{i}sica, CINVESTAV for hospitality. G.L.A. thanks Brown University
HET Group for kind hospitality and B. Franchi and S. Graffi for their
help. This research is supported in part by DOE grant DE-FE0291ER40688-Task
A. D.L. and K.G. are supported in part by US--Israel Bi-national Science
Foundation grant \#2000359.

\bibliographystyle{apsrev}
\clearpage\addcontentsline{toc}{chapter}{\bibname}\bibliography{desitterqft}

\begin{thebibliography}{42}
\expandafter\ifx\csname natexlab\endcsname\relax\def\natexlab#1{#1}\fi
\expandafter\ifx\csname bibnamefont\endcsname\relax
  \def\bibnamefont#1{#1}\fi
\expandafter\ifx\csname bibfnamefont\endcsname\relax
  \def\bibfnamefont#1{#1}\fi
\expandafter\ifx\csname citenamefont\endcsname\relax
  \def\citenamefont#1{#1}\fi
\expandafter\ifx\csname url\endcsname\relax
  \def\url#1{\texttt{#1}}\fi
\expandafter\ifx\csname urlprefix\endcsname\relax\def\urlprefix{URL }\fi
\providecommand{\bibinfo}[2]{#2}
\providecommand{\eprint}[2][]{\url{#2}}

\bibitem[{\citenamefont{Chernikov and Tagirov}(1968)}]{Chernikov:1968zm}
\bibinfo{author}{\bibfnamefont{N.~A.} \bibnamefont{Chernikov}}
  \bibnamefont{and} \bibinfo{author}{\bibfnamefont{E.~A.}
  \bibnamefont{Tagirov}}, \bibinfo{journal}{Annales Poincare Phys. Theor.}
  \textbf{\bibinfo{volume}{A9}}, \bibinfo{pages}{109} (\bibinfo{year}{1968}).

\bibitem[{\citenamefont{Tagirov}(1973)}]{Tagirov:1973vv}
\bibinfo{author}{\bibfnamefont{E.~A.} \bibnamefont{Tagirov}},
  \bibinfo{journal}{Ann. Phys.} \textbf{\bibinfo{volume}{76}},
  \bibinfo{pages}{561} (\bibinfo{year}{1973}).

\bibitem[{\citenamefont{Allen}(1985)}]{Allen:1985ux}
\bibinfo{author}{\bibfnamefont{B.}~\bibnamefont{Allen}},
  \bibinfo{journal}{Phys. Rev.} \textbf{\bibinfo{volume}{D32}},
  \bibinfo{pages}{3136} (\bibinfo{year}{1985}).

\bibitem[{\citenamefont{Mottola}(1985)}]{Mottola:1985ar}
\bibinfo{author}{\bibfnamefont{E.}~\bibnamefont{Mottola}},
  \bibinfo{journal}{Phys. Rev.} \textbf{\bibinfo{volume}{D31}},
  \bibinfo{pages}{754} (\bibinfo{year}{1985}).

\bibitem[{\citenamefont{Easther et~al.}(2001)\citenamefont{Easther, Greene,
  Kinney, and Shiu}}]{Easther:2001fi}
\bibinfo{author}{\bibfnamefont{R.}~\bibnamefont{Easther}},
  \bibinfo{author}{\bibfnamefont{B.~R.} \bibnamefont{Greene}},
  \bibinfo{author}{\bibfnamefont{W.~H.} \bibnamefont{Kinney}},
  \bibnamefont{and} \bibinfo{author}{\bibfnamefont{G.}~\bibnamefont{Shiu}},
  \bibinfo{journal}{Phys. Rev.} \textbf{\bibinfo{volume}{D64}},
  \bibinfo{pages}{103502} (\bibinfo{year}{2001}), \eprint{hep-th/0104102}.

\bibitem[{\citenamefont{Easther et~al.}(2003)\citenamefont{Easther, Greene,
  Kinney, and Shiu}}]{Easther:2001fz}
\bibinfo{author}{\bibfnamefont{R.}~\bibnamefont{Easther}},
  \bibinfo{author}{\bibfnamefont{B.~R.} \bibnamefont{Greene}},
  \bibinfo{author}{\bibfnamefont{W.~H.} \bibnamefont{Kinney}},
  \bibnamefont{and} \bibinfo{author}{\bibfnamefont{G.}~\bibnamefont{Shiu}},
  \bibinfo{journal}{Phys. Rev.} \textbf{\bibinfo{volume}{D67}},
  \bibinfo{pages}{063508} (\bibinfo{year}{2003}), \eprint{hep-th/0110226}.

\bibitem[{\citenamefont{Easther et~al.}(2002)\citenamefont{Easther, Greene,
  Kinney, and Shiu}}]{Easther:2002xe}
\bibinfo{author}{\bibfnamefont{R.}~\bibnamefont{Easther}},
  \bibinfo{author}{\bibfnamefont{B.~R.} \bibnamefont{Greene}},
  \bibinfo{author}{\bibfnamefont{W.~H.} \bibnamefont{Kinney}},
  \bibnamefont{and} \bibinfo{author}{\bibfnamefont{G.}~\bibnamefont{Shiu}},
  \bibinfo{journal}{Phys. Rev.} \textbf{\bibinfo{volume}{D66}},
  \bibinfo{pages}{023518} (\bibinfo{year}{2002}), \eprint{hep-th/0204129}.

\bibitem[{\citenamefont{Bergstrom and Danielsson}(2002)}]{Bergstrom:2002yd}
\bibinfo{author}{\bibfnamefont{L.}~\bibnamefont{Bergstrom}} \bibnamefont{and}
  \bibinfo{author}{\bibfnamefont{U.~H.} \bibnamefont{Danielsson}},
  \bibinfo{journal}{JHEP} \textbf{\bibinfo{volume}{12}}, \bibinfo{pages}{038}
  (\bibinfo{year}{2002}), \eprint{hep-th/0211006}.

\bibitem[{\citenamefont{Danielsson}(2002{\natexlab{a}})}]{Danielsson:2002kx}
\bibinfo{author}{\bibfnamefont{U.~H.} \bibnamefont{Danielsson}},
  \bibinfo{journal}{Phys. Rev.} \textbf{\bibinfo{volume}{D66}},
  \bibinfo{pages}{023511} (\bibinfo{year}{2002}{\natexlab{a}}),
  \eprint{hep-th/0203198}.

\bibitem[{\citenamefont{Danielsson}(2002{\natexlab{b}})}]{Danielsson:2002mb}
\bibinfo{author}{\bibfnamefont{U.~H.} \bibnamefont{Danielsson}},
  \bibinfo{journal}{JHEP} \textbf{\bibinfo{volume}{12}}, \bibinfo{pages}{025}
  (\bibinfo{year}{2002}{\natexlab{b}}), \eprint{hep-th/0210058}.

\bibitem[{\citenamefont{Danielsson}(2002{\natexlab{c}})}]{Danielsson:2002qh}
\bibinfo{author}{\bibfnamefont{U.~H.} \bibnamefont{Danielsson}},
  \bibinfo{journal}{JHEP} \textbf{\bibinfo{volume}{07}}, \bibinfo{pages}{040}
  (\bibinfo{year}{2002}{\natexlab{c}}), \eprint{hep-th/0205227}.

\bibitem[{\citenamefont{Goldstein and
  Lowe}(2003{\natexlab{a}})}]{Goldstein:2002fc}
\bibinfo{author}{\bibfnamefont{K.}~\bibnamefont{Goldstein}} \bibnamefont{and}
  \bibinfo{author}{\bibfnamefont{D.~A.} \bibnamefont{Lowe}},
  \bibinfo{journal}{Phys. Rev.} \textbf{\bibinfo{volume}{D67}},
  \bibinfo{pages}{063502} (\bibinfo{year}{2003}{\natexlab{a}}),
  \eprint{hep-th/0208167}.

\bibitem[{\citenamefont{Alberghi et~al.}(2003)\citenamefont{Alberghi, Casadio,
  and Tronconi}}]{Alberghi:2003am}
\bibinfo{author}{\bibfnamefont{G.~L.} \bibnamefont{Alberghi}},
  \bibinfo{author}{\bibfnamefont{R.}~\bibnamefont{Casadio}}, \bibnamefont{and}
  \bibinfo{author}{\bibfnamefont{A.}~\bibnamefont{Tronconi}}
  (\bibinfo{year}{2003}), \eprint{gr-qc/0303035}.

\bibitem[{\citenamefont{Starobinsky and Tkachev}(2002)}]{Starobinsky:2002rp}
\bibinfo{author}{\bibfnamefont{A.~A.} \bibnamefont{Starobinsky}}
  \bibnamefont{and} \bibinfo{author}{\bibfnamefont{I.~I.}
  \bibnamefont{Tkachev}}, \bibinfo{journal}{JETP Lett.}
  \textbf{\bibinfo{volume}{76}}, \bibinfo{pages}{235} (\bibinfo{year}{2002}),
  \eprint{astro-ph/0207572}.

\bibitem[{\citenamefont{Anchordoqui et~al.}(2003)\citenamefont{Anchordoqui,
  Paul, Reucroft, and Swain}}]{Anchordoqui}
\bibinfo{author}{\bibfnamefont{L.}~\bibnamefont{Anchordoqui}},
  \bibinfo{author}{\bibfnamefont{T.}~\bibnamefont{Paul}},
  \bibinfo{author}{\bibfnamefont{S.}~\bibnamefont{Reucroft}}, \bibnamefont{and}
  \bibinfo{author}{\bibfnamefont{J.}~\bibnamefont{Swain}},
  \bibinfo{journal}{Int. J. Mod. Phys.} \textbf{\bibinfo{volume}{A18}},
  \bibinfo{pages}{2229} (\bibinfo{year}{2003}), \eprint{hep-ph/0206072}.

\bibitem[{\citenamefont{Bhattacharjee and Sigl}(2000)}]{Bhattacharjee:1998qc}
\bibinfo{author}{\bibfnamefont{P.}~\bibnamefont{Bhattacharjee}}
  \bibnamefont{and} \bibinfo{author}{\bibfnamefont{G.}~\bibnamefont{Sigl}},
  \bibinfo{journal}{Phys. Rept.} \textbf{\bibinfo{volume}{327}},
  \bibinfo{pages}{109} (\bibinfo{year}{2000}), \eprint{astro-ph/9811011}.

\bibitem[{\citenamefont{Greisen}(1966)}]{Greisen:1966jv}
\bibinfo{author}{\bibfnamefont{K.}~\bibnamefont{Greisen}},
  \bibinfo{journal}{Phys. Rev. Lett.} \textbf{\bibinfo{volume}{16}},
  \bibinfo{pages}{748} (\bibinfo{year}{1966}).

\bibitem[{\citenamefont{Zatsepin and Kuzmin}(1966)}]{Zatsepin:1966jv}
\bibinfo{author}{\bibfnamefont{G.~T.} \bibnamefont{Zatsepin}} \bibnamefont{and}
  \bibinfo{author}{\bibfnamefont{V.~A.} \bibnamefont{Kuzmin}},
  \bibinfo{journal}{JETP Lett.} \textbf{\bibinfo{volume}{4}},
  \bibinfo{pages}{78} (\bibinfo{year}{1966}).

\bibitem[{\citenamefont{Takeda et~al.}(1998)}]{Takeda:1998ps}
\bibinfo{author}{\bibfnamefont{M.}~\bibnamefont{Takeda}} \bibnamefont{et~al.},
  \bibinfo{journal}{Phys. Rev. Lett.} \textbf{\bibinfo{volume}{81}},
  \bibinfo{pages}{1163} (\bibinfo{year}{1998}), \eprint{astro-ph/9807193}.

\bibitem[{\citenamefont{Hayashida et~al.}(1999)}]{Hayashida:2000zr}
\bibinfo{author}{\bibfnamefont{N.}~\bibnamefont{Hayashida}}
  \bibnamefont{et~al.}, \bibinfo{journal}{Astrophys. J.}
  \textbf{\bibinfo{volume}{522}}, \bibinfo{pages}{225} (\bibinfo{year}{1999}),
  \eprint{astro-ph/0008102}.

\bibitem[{\citenamefont{Abu-Zayyad et~al.}(2002)}]{Hires}
\bibinfo{author}{\bibfnamefont{T.}~\bibnamefont{Abu-Zayyad}}
  \bibnamefont{et~al.} (\bibinfo{collaboration}{High Resolution Fly's Eye})
  (\bibinfo{year}{2002}), \eprint{astro-ph/0208243}.

\bibitem[{\citenamefont{Bousso et~al.}(2002)\citenamefont{Bousso, Maloney, and
  Strominger}}]{Bousso:2001mw}
\bibinfo{author}{\bibfnamefont{R.}~\bibnamefont{Bousso}},
  \bibinfo{author}{\bibfnamefont{A.}~\bibnamefont{Maloney}}, \bibnamefont{and}
  \bibinfo{author}{\bibfnamefont{A.}~\bibnamefont{Strominger}},
  \bibinfo{journal}{Phys. Rev.} \textbf{\bibinfo{volume}{D65}},
  \bibinfo{pages}{104039} (\bibinfo{year}{2002}), \eprint{hep-th/0112218}.

\bibitem[{\citenamefont{Kuzmin and Rubakov}(1998)}]{Kuzmin:1998cm}
\bibinfo{author}{\bibfnamefont{V.~A.} \bibnamefont{Kuzmin}} \bibnamefont{and}
  \bibinfo{author}{\bibfnamefont{V.~A.} \bibnamefont{Rubakov}},
  \bibinfo{journal}{Phys. Atom. Nucl.} \textbf{\bibinfo{volume}{61}},
  \bibinfo{pages}{1028} (\bibinfo{year}{1998}), \eprint{astro-ph/9709187}.

\bibitem[{\citenamefont{Birkel and Sarkar}(1998)}]{Birkel:1998nx}
\bibinfo{author}{\bibfnamefont{M.}~\bibnamefont{Birkel}} \bibnamefont{and}
  \bibinfo{author}{\bibfnamefont{S.}~\bibnamefont{Sarkar}},
  \bibinfo{journal}{Astropart. Phys.} \textbf{\bibinfo{volume}{9}},
  \bibinfo{pages}{297} (\bibinfo{year}{1998}), \eprint{hep-ph/9804285}.

\bibitem[{\citenamefont{Berezinsky et~al.}(1998)\citenamefont{Berezinsky,
  Blasi, and Vilenkin}}]{Berezinsky:1998qv}
\bibinfo{author}{\bibfnamefont{V.}~\bibnamefont{Berezinsky}},
  \bibinfo{author}{\bibfnamefont{P.}~\bibnamefont{Blasi}}, \bibnamefont{and}
  \bibinfo{author}{\bibfnamefont{A.}~\bibnamefont{Vilenkin}},
  \bibinfo{journal}{Phys. Rev.} \textbf{\bibinfo{volume}{D58}},
  \bibinfo{pages}{103515} (\bibinfo{year}{1998}).

\bibitem[{\citenamefont{Goldstein and
  Lowe}(2003{\natexlab{b}})}]{Goldstein:2003ut}
\bibinfo{author}{\bibfnamefont{K.}~\bibnamefont{Goldstein}} \bibnamefont{and}
  \bibinfo{author}{\bibfnamefont{D.~A.} \bibnamefont{Lowe}},
  \bibinfo{journal}{accepted for publication, Nucl. Phys. B}
  (\bibinfo{year}{2003}{\natexlab{b}}), \eprint{hep-th/0302050}.

\bibitem[{\citenamefont{Birrell and Davies}(1982)}]{Birrell:1982ix}
\bibinfo{author}{\bibfnamefont{N.~D.} \bibnamefont{Birrell}} \bibnamefont{and}
  \bibinfo{author}{\bibfnamefont{P.~C.~W.} \bibnamefont{Davies}}
  (\bibinfo{year}{1982}), \bibinfo{note}{cambridge, Uk: Univ. Pr. 340p}.

\bibitem[{\citenamefont{Martin and Brandenberger}(2001)}]{Martin:2000xs}
\bibinfo{author}{\bibfnamefont{J.}~\bibnamefont{Martin}} \bibnamefont{and}
  \bibinfo{author}{\bibfnamefont{R.~H.} \bibnamefont{Brandenberger}},
  \bibinfo{journal}{Phys. Rev.} \textbf{\bibinfo{volume}{D63}},
  \bibinfo{pages}{123501} (\bibinfo{year}{2001}), \eprint{hep-th/0005209}.

\bibitem[{\citenamefont{Greenberg}(2002)}]{Greenberg:2002uu}
\bibinfo{author}{\bibfnamefont{O.~W.} \bibnamefont{Greenberg}},
  \bibinfo{journal}{Phys. Rev. Lett.} \textbf{\bibinfo{volume}{89}},
  \bibinfo{pages}{231602} (\bibinfo{year}{2002}), \eprint{hep-ph/0201258}.

\bibitem[{\citenamefont{Barbot and Drees}(2002)}]{Barbot:2002gt}
\bibinfo{author}{\bibfnamefont{C.}~\bibnamefont{Barbot}} \bibnamefont{and}
  \bibinfo{author}{\bibfnamefont{M.}~\bibnamefont{Drees}}
  (\bibinfo{year}{2002}), \eprint{hep-ph/0211406}.

\bibitem[{\citenamefont{Sigl}(2003)}]{Sigl:2002yk}
\bibinfo{author}{\bibfnamefont{G.}~\bibnamefont{Sigl}}, \bibinfo{journal}{Ann.
  Phys.} \textbf{\bibinfo{volume}{303}}, \bibinfo{pages}{117}
  (\bibinfo{year}{2003}), \eprint{astro-ph/0210049}.

\bibitem[{\citenamefont{Ave et~al.}(2000)\citenamefont{Ave, Hinton, Vazquez,
  Watson, and Zas}}]{Ave:2000nd}
\bibinfo{author}{\bibfnamefont{M.}~\bibnamefont{Ave}},
  \bibinfo{author}{\bibfnamefont{J.~A.} \bibnamefont{Hinton}},
  \bibinfo{author}{\bibfnamefont{R.~A.} \bibnamefont{Vazquez}},
  \bibinfo{author}{\bibfnamefont{A.~A.} \bibnamefont{Watson}},
  \bibnamefont{and} \bibinfo{author}{\bibfnamefont{E.}~\bibnamefont{Zas}},
  \bibinfo{journal}{Phys. Rev. Lett.} \textbf{\bibinfo{volume}{85}},
  \bibinfo{pages}{2244} (\bibinfo{year}{2000}), \eprint{astro-ph/0007386}.

\bibitem[{\citenamefont{Ave et~al.}(2002)\citenamefont{Ave, Hinton, Vazquez,
  Watson, and Zas}}]{Ave:2001xn}
\bibinfo{author}{\bibfnamefont{M.}~\bibnamefont{Ave}},
  \bibinfo{author}{\bibfnamefont{J.~A.} \bibnamefont{Hinton}},
  \bibinfo{author}{\bibfnamefont{R.~A.} \bibnamefont{Vazquez}},
  \bibinfo{author}{\bibfnamefont{A.~A.} \bibnamefont{Watson}},
  \bibnamefont{and} \bibinfo{author}{\bibfnamefont{E.}~\bibnamefont{Zas}},
  \bibinfo{journal}{Phys. Rev.} \textbf{\bibinfo{volume}{D65}},
  \bibinfo{pages}{063007} (\bibinfo{year}{2002}), \eprint{astro-ph/0110613}.

\bibitem[{\citenamefont{Shinozaki et~al.}(2002)}]{Shinozaki:2002ve}
\bibinfo{author}{\bibfnamefont{K.}~\bibnamefont{Shinozaki}}
  \bibnamefont{et~al.}, \bibinfo{journal}{Astrophys. J.}
  \textbf{\bibinfo{volume}{571}}, \bibinfo{pages}{L117} (\bibinfo{year}{2002}).

\bibitem[{\citenamefont{Cronin}(1992)}]{Cronin:1992ir}
\bibinfo{author}{\bibfnamefont{J.~W.} \bibnamefont{Cronin}},
  \bibinfo{journal}{Nucl. Phys. Proc. Suppl.} \textbf{\bibinfo{volume}{28B}},
  \bibinfo{pages}{213} (\bibinfo{year}{1992}).

\bibitem[{\citenamefont{Kaloper et~al.}(2002)\citenamefont{Kaloper, Kleban,
  Lawrence, Shenker, and Susskind}}]{Kaloper:2002cs}
\bibinfo{author}{\bibfnamefont{N.}~\bibnamefont{Kaloper}},
  \bibinfo{author}{\bibfnamefont{M.}~\bibnamefont{Kleban}},
  \bibinfo{author}{\bibfnamefont{A.}~\bibnamefont{Lawrence}},
  \bibinfo{author}{\bibfnamefont{S.}~\bibnamefont{Shenker}}, \bibnamefont{and}
  \bibinfo{author}{\bibfnamefont{L.}~\bibnamefont{Susskind}},
  \bibinfo{journal}{JHEP} \textbf{\bibinfo{volume}{11}}, \bibinfo{pages}{037}
  (\bibinfo{year}{2002}), \eprint{hep-th/0209231}.

\bibitem[{\citenamefont{Banks and Mannelli}(2002)}]{Banks:2002nv}
\bibinfo{author}{\bibfnamefont{T.}~\bibnamefont{Banks}} \bibnamefont{and}
  \bibinfo{author}{\bibfnamefont{L.}~\bibnamefont{Mannelli}}
  (\bibinfo{year}{2002}), \eprint{hep-th/0209113}.

\bibitem[{\citenamefont{Einhorn and Larsen}(2002)}]{Einhorn:2002nu}
\bibinfo{author}{\bibfnamefont{M.~B.} \bibnamefont{Einhorn}} \bibnamefont{and}
  \bibinfo{author}{\bibfnamefont{F.}~\bibnamefont{Larsen}}
  (\bibinfo{year}{2002}), \eprint{hep-th/0209159}.

\bibitem[{\citenamefont{Einhorn and Larsen}(2003)}]{Einhorn:2003xb}
\bibinfo{author}{\bibfnamefont{M.~B.} \bibnamefont{Einhorn}} \bibnamefont{and}
  \bibinfo{author}{\bibfnamefont{F.}~\bibnamefont{Larsen}}
  (\bibinfo{year}{2003}), \eprint{hep-th/0305056}.

\bibitem[{\citenamefont{Collins et~al.}(2003)\citenamefont{Collins, Holman, and
  Martin}}]{Collins:2003zv}
\bibinfo{author}{\bibfnamefont{H.}~\bibnamefont{Collins}},
  \bibinfo{author}{\bibfnamefont{R.}~\bibnamefont{Holman}}, \bibnamefont{and}
  \bibinfo{author}{\bibfnamefont{M.~R.} \bibnamefont{Martin}}
  (\bibinfo{year}{2003}), \eprint{hep-th/0306028}.

\bibitem[{\citenamefont{Goldstein and Lowe}(2003{\natexlab{c}})}]{lowe3}
\bibinfo{author}{\bibfnamefont{K.}~\bibnamefont{Goldstein}} \bibnamefont{and}
  \bibinfo{author}{\bibfnamefont{D.~A.} \bibnamefont{Lowe}}
  (\bibinfo{year}{2003}{\natexlab{c}}), \eprint{hep-th/0308135}.

\bibitem[{\citenamefont{Parikh et~al.}(2003)\citenamefont{Parikh, Savonije, and
  Verlinde}}]{Parikh:2002py}
\bibinfo{author}{\bibfnamefont{M.~K.} \bibnamefont{Parikh}},
  \bibinfo{author}{\bibfnamefont{I.}~\bibnamefont{Savonije}}, \bibnamefont{and}
  \bibinfo{author}{\bibfnamefont{E.}~\bibnamefont{Verlinde}},
  \bibinfo{journal}{Phys. Rev.} \textbf{\bibinfo{volume}{D67}},
  \bibinfo{pages}{064005} (\bibinfo{year}{2003}), \eprint{hep-th/0209120}.

\end{thebibliography}
\end{acknowledgments}

\end{document}